\documentclass[%
superscriptaddress,
preprint,
amsmath,amssymb,
prb,
]{revtex4-2}

\usepackage{graphicx}% Include figure files
\usepackage{dcolumn}% Align table columns on decimal point
\usepackage{bm}% bold math
\usepackage{float}% bold math
\begin{document}

\preprint{}

\title{First-principles analysis of energy exchange in time-varying capacitors for energy trapping applications}% Force line breaks with \\
\author{Kristy Hecht}
\affiliation{Department of Electrical and Computer Engineering, University of North Carolina at Charlotte, Charlotte, North Carolina 28223, USA}
\author{David Gonz\'alez-Ovejero}
\affiliation{CNRS, Institut d'\'Electronique et des Technologies du num\'eRique (IETR), UMR 6164, University of Rennes, F-35000 Rennes, France}
\author{Dimitrios L. Sounas}
\affiliation{Department of Electrical and Computer Engineering, Wayne State University, Detroit, Michigan 48202, USA}
\author{Mario Junior Mencagli}
\thanks{Corresponding author: mmencagl@uncc.edu}
%\email[]{mmencagl@uncc.edu}
\affiliation{Department of Electrical and Computer Engineering, University of North Carolina at Charlotte, Charlotte, North Carolina 28223, USA}


%\affiliation{
%\affiliation{
%\affiliation{%


\date{\today}% It is always \today, today,

\begin{abstract}
Time-varying networks, consisting of lumped elements, such as resistors, capacitors, and inductors, actively modulated in time, have introduced a host of novel wave phenomena and witnessed a remarkable development during recent years. This paper investigates the scattering from a time-varying capacitor and how such a load can be fully reflectionless when the capacitance is suitably modulated in time. We analytically derive the required temporal dependence of the capacitance and show how in contrast to other techniques it avoids extreme and negative values and, as a result, can be implemented in a feasible way, when the capacitor is charged with a DC voltage source. We also derive from first principles the energy balance of such a time-varying capacitor, proving that the energy of an incoming pulse is transferred to the modulation source. Our findings clarify scattering of waves from time-varying capacitors and open up a new way to matching of broadband pulses.
\end{abstract}

\maketitle

%\tableofcontents

%\section{\label{sec:level1}First-level heading:\protect\\ The line
%break was forced \lowercase{via} \textbackslash\textbackslash}
%%%%%%%%%%%%%%%%%%%%%%%%%%%%%%%%%%%%%%%%%%%%%%%%%%%%%%%%%
\section{Introduction} \label{sec1}
%%%%%%%%%%%%%%%%%%%%%%%%%%%%%%%%%%%%%%%%%%%%%%%%%%%%%%%%%
One of the hottest topics in current electromagnetics and photonics research is the study of structures with parameters, such as permittivity and/or permeability, varying in time.  Exploiting time as a new degree of freedom for the control of electromagnetic waves has enabled the development of structures with intriguing functionalities, such as time-Floquet topological insulators \cite{Lindner2011,Katan2013,Lustig18}, temporal-based non-reciprocity \cite{Yu2009,Estep2014,Sounas2017}, static-to-dynamic field conversion \cite{mencagli2021}, that overcome most of the challenges faced by time-invariant structures. Recently, the platform of time-varying media has also been combined with the concept of metamaterials, opening up another interesting avenue to control and achieve desired functionalities in wave-matter interaction \cite{Amir2018,Ptitcyn2019,Caloz2020,Pacheco2020,Pacheco2021,Wang2021,Wu2020,Engheta2021,Huidobro2021}. In addition to physics and engineering, research on time-variant systems has been aggressively pursued in mechanics \cite{Darabi2020}, acoustics \cite{Fleury2016}, and water-wave \cite{Bacot2016}.

An important subclass of electromagnetic time-varying media is represented by time-varying networks consisting of lumped elements, such as resistors, capacitors, and inductors with time-dependent properties. These networks have stimulated a great deal of research interest \cite{Kamal1960,Anderson1965,Cullen1960,Cassedy1965,Estep2014,Kord2018,Wang2020}. In addition to providing an interesting platform to control and manipulate electromagnetic waves, they are also more amenable for experimental demonstrations than the conventional time-varying media \cite{Estep2014,Kord2018,Wang2020}. The vast majority of studies related to temporal-dependent networks have been limited to lumped elements periodically modulated in time. On the other hand, lumped elements with aperiodic time modulation can provide an extra degree of freedom to engineer time-variant networks with new functionalities. Indeed, it has been recently shown how reactive elements, like capacitors and inductors, judiciously modulated in time can provide an alternative for electromagnetic energy accumulation in a transmission line scenario \cite{Mirmoosa2019,Sounas2020}. For example, \cite{Mirmoosa2019} showed how reactive elements, albeit lossless, when experiencing a temporal variation, exhibit a resistive behavior responsible for the incoming signal absorption. However, this approach requires capacitors or inductors with extreme values, including negative ones, which may be challenging to achieve from a practical standpoint. A follow-up proposal \cite{Sounas2020} showed that this issue can be overcome if modulation is applied to a coupling network between an LC-resonator and a feeding transmission line, which allows transferring the incoming energy to the LC network by applying slow modulation. Both of these approaches work for harmonic input signals and no technique exists for capturing of arbitrary pulses.

The work is motivated by the recent interest on electromagnetic energy accumulation enabled by time-varying lumped elements. We propose here an approach for capturing the energy of an arbitrary pulse by applying a non-periodic modulation to a single capacitor at the end of a transmission line. We elucidate why the previously proposed approach \cite{Mirmoosa2019} requires extreme capacitance values and how combining an incoming pulse with a DC signal allows one to overcome such an issue. Furthermore, to gain further insight, we derive from the first principles the system's energy balance, proving that the energy of the pulse is actually transferred to the modulation source, instead of being stored in the capacitor. 

The paper is organized as follows. Sec.~\ref{sec2} introduces the system under investigation consisting of a transmission line terminated with a time-varying capacitor that is charged up by a DC voltage connected at the other end of the transmission line. In Sec.~\ref{sec3}, we derive the expression of the required temporal profile of the capacitance to make the system reflectionless. This expression is then further investigated by considering two illustrative incoming voltage pulses in Secs.~\ref{subsec3A} and \ref{subsec3B}. In Sec.~\ref{sec4}, we derive from first principles the energy balance of the system. Secs.~\ref{subsec4A} and \ref{subsec4B} describes the energy exchange mechanism for the incoming pulses of Secs.~\ref{subsec3A} and \ref{subsec3B}, respectively. Finally, we draw our conclusions in Sec.~\ref{sec5}.

%%%%%%%%%%%%%%%%%%%%%%%%%%%%%%%%%%%%%%%%%%%%%%%%%%%%%%%%%
\section{Statement of the problem} \label{sec2}
\begin{figure}[ht]
\centering
\includegraphics[width=0.6\columnwidth]{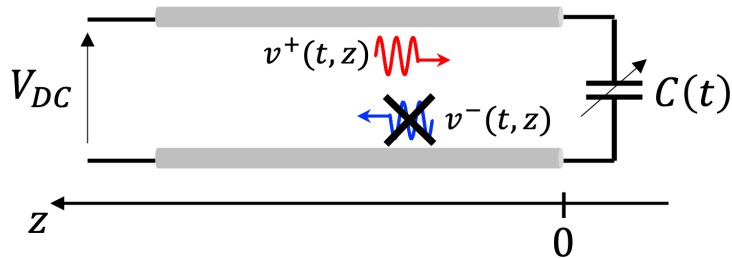}
\caption{A DC voltage source $V_{dc}$ is connected to a time-varying capacitor through a transmission line with characteristic impedance $Z_0$. The incoming voltage pulse $v^+(t,z)$ (red arrow) travels toward the capacitor, while the reflected voltage pulse $v^-(t,z)$ (blue arrow) travels away from the same capacitive load. The capacitor can eliminate the reflected pulse as long as its capacitance experiences a suitable temporal modulation.} 
\label{one}
\end{figure}

%%%%%%%%%%%%%%%%%%%%%%%%%%%%%%%%%%%%%%%%%%%%%%%%%%%%%%%%%
A sketch of the system under study is depicted in Fig.~\ref{one}: a lossless transmission line is terminated with a time-varying capacitor. A DC voltage source ($V_{dc}$) is connected to the input port. Once the DC source has fully charged up the capacitor, say at $t=-\infty$, a voltage pulse [${v^ + }\left( {t,z} \right)$] starts propagating along the transmission line toward the capacitive load. We assume that the DC voltage source continues being connected to the system even for $t>-\infty$. Also, without loss of generality and for the sake of simplicity, we assume $V_{dc}>0$ throughout the paper. Denoting by ${v^ - }\left( {t,z} \right)$ the reflected voltage pulse, the total instantaneous voltage and current at the load location ($z=0$) can be expressed as 
%%%%%%%%%%%%%%%%%%%%  v_i_load  %%%%%%%%%%%%%%%%%%%%%%%%%%%%%%%
\begin{subequations}\label{v_i_load}
\begin{equation}\label{eq1a}
\begin{aligned}
v\left( t \right) = {v^ + }\left( {t,0} \right) + {v^ - }\left( {t,0} \right) + {V_{dc}}
\end{aligned}
\end{equation}
\begin{equation}\label{eq1b}
\begin{aligned}
i\left( t \right) = \frac{{{v^ + }\left( {t,0} \right) - {v^ - }\left( {t,0} \right)}}{{{Z_0}}}
\end{aligned}
\end{equation}
\end{subequations}
%%%%%%%%%%%%%%%%%%%%%%%%%%%%%%%%%%%%%%%%%%%%%%%%%%%%%%%%%
with $Z_0$ being the characteristic impedance of the transmission line. Now, to discuss the possibility of trapping the energy of the incoming voltage pulse in the capacitor, the first step is to find the temporal variation of its capacitance to fully eliminate the reflected pulse, which is the subject of the next section.
%%%%%%%%%%%%%%%%%%%%  Fig. 1   %%%%%%%%%%%%%%%%%%%%%%%%%%%%%%%%%
%%%%%%%%%%%%%%%%%%%%%%%%%%%%%%%%%%%%%%%%%%%%%%%%%%%%%%%%%

%%%%%%%%%%%%%%%%%%%%%%%%%%%%%%%%%%%%%%%%%%%%%%%%%%%%%%%%%
\section{Reflectionless time-dependent capacitor} \label{sec3}
%%%%%%%%%%%%%%%%%%%%%%%%%%%%%%%%%%%%%%%%%%%%%%%%%%%%%%%%%
It is very well-known that a stationary (no time-varying) purely capacitive load presents a reflection coefficient with a magnitude equal to one. Therefore, an incident pulse, upon reaching this load, will be totally reflected back. Here, leveraging the extra degree of freedom provided by the temporal dependence of the capacitance, we show that the reflected pulse can be fully eliminated. To this end, first, we impose zero reflection (${v^ - }\left( {t,z} \right)=0$), and then substituting Eqs.~(\ref{v_i_load}a) and (\ref{v_i_load}b) into the $i-v$ capacitor equation ($i = \frac{d}{{dt}}\left( {Cv} \right)$) we obtain the following first-order ordinary differential equation
%%%%%%%%%%%%%%%%%%%%  ode_ct  %%%%%%%%%%%%%%%%%%%%%%%%%%%%%%%%
\begin{equation}\label{ode_ct}
\frac{{dC\left( t \right)}}{dt}+\frac{{C\left( t \right)}}{{v^ + }\left( {t,0} \right)+V_{dc}}\frac{d{v^ + }\left( {t,0} \right)}{dt}=\frac{{v^ + }\left( {t,0} \right)}{Z_0\left({v^ + }\left( {t,0} \right)+V_{dc}\right)}
\end{equation}
%%%%%%%%%%%%%%%%%%%%%%%%%%%%%%%%%%%%%%%%%%%%%%%%%%%%%%%%%
with a solution with respect to $C\left( {t} \right)$ as
%%%%%%%%%%%%%%%%%%%%  ct_refless  %%%%%%%%%%%%%%%%%%%%%%%%%%%%%%%
\begin{equation}\label{ct_refless}
{C}\left( t \right) = \frac{{{v^ + }\left( {-\infty,0} \right) + {V_{dc}}}}{{{v^ + }\left( {t,0} \right) + {V_{dc}}}}\left[ {\frac{1}{{{Z_0}\left( {{v^ + }\left( {-\infty,0} \right) + {V_{dc}}} \right)}}} 
{\int\limits_{{-\infty}}^t {{v^ + }\left( {\varepsilon ,0} \right)d\varepsilon }  + C_i} \right]
\end{equation}
%%%%%%%%%%%%%%%%%%%%%%%%%%%%%%%%%%%%%%%%%%%%%%%%%%%%%%%%%
This equation, given the incoming pulse ${v^ + }\left( {t,z} \right)$, provides the required temporal variation of the capacitance with initial value $C_i$ to eliminate the reflected pulse in the system of Fig.~\ref{one}. Inspecting Eq.~(\ref{ct_refless}), one can observe that ${C}\left( t \right)$ diverges for ${{v^ + }\left( {t,0} \right) =- {V_{dc}}}$. In order to avoid extreme capacitive values, it is important to have a DC voltage across the capacitor outside the range of values of the incoming voltage pulse (${V_{dc}}>|{v^ + }\left( {t,0} \right)|$). This simple analysis elucidates the extreme capacitive values required in the approach proposed in \cite{Mirmoosa2019}, which did not include a DC source and, as a result, required a capacitive value tending to infinity. 

Another important physical quantity to bring into the discussion, as will become clear shortly, is the temporal variation of the charge in the capacitor to cancel the reflected pulse. Recalling the capacitor charge-voltage relationship ($q=Cv$), it is straightforward to derive the temporal variation of the charge in the capacitor with respect to its initial value ($q_i={C_i}{V_{dc}}$) from Eq.~(\ref{ct_refless})
%%%%%%%%%%%%%%%%%%%%  dqt  %%%%%%%%%%%%%%%%%%%%%%%%%%%%%%%%%%
\begin{equation}\label{dqt}
\Delta {q}\left( t \right) = {{C}\left( t \right)} \left[ {{V_{dc}} + v^ + \left( t ,0\right)} \right] - C_{i} V_{dc}
\end{equation}
%%%%%%%%%%%%%%%%%%%%%%%%%%%%%%%%%%%%%%%%%%%%%%%%%%%%%%%%%
To further investigate the above two equations and their principal physical insights, now we consider two different temporal shapes of incoming voltage pulses.
%%%%%%%%%%%%%%%%%%%%%%%%%%%%%%%%%%%%%%%%%%%%%%%%%%%%%%%%%%%%%%%%%%%%%%%%%%%%%%%%%%%%%%%%%%
\subsection{First-order derivative of a Gaussian pulse} \label{subsec3A}
%%%%%%%%%%%%%%%%%%%%%%%%%%%%%%%%%%%%%%%%%%%%%%%%%%%%%%%%%%%%%%%%%%%%%%%%%%%%%%%%%%%%%%%%%%
The mathematical expression of a normalized voltage pulse with a temporal profile of the first-order derivative of the Gaussian function, which is shown in the inset of  Fig.~\ref{two}(a), can be written as  $v_{dg}^ + \left( {t,z} \right) =  - \frac{{t - {v_p}z - \mu }}{\sigma^2 }{e^{ - \frac{{{{\left( {t - {v_p}z - \mu } \right)}^2} - {\sigma ^2}}}{{2{\sigma ^2}}}}}$, with $\mu$ being the zero crossing position, $v_p$ the phase velocity, and $\sigma$ a positive real number controlling the width of the pulse. Substituting $v_{dg}^ + \left( {t,z} \right)$ into Eq.~(\ref{ct_refless}) and simplifying, we get
%As first incoming pulse, we consider the first-order derivative of a gaussian pulse (see inset of Fig.~\ref{two}(a)).  Its mathematical expression reads $v_{dg}^ + \left( {t,z} \right) =  - \frac{{t - {v_p}z - \mu }}{\sigma }{e^{ - \frac{{{{\left( {t - {v_p}z - \mu } \right)}^2} - {\sigma ^2}}}{{2{\sigma ^2}}}}}$ with $\mu$ the location of the zero crossing, $\sigma$ the standard deviation, and $v_p$ the phase velocity. Plugging $v_{dg}^ + \left( {t,z} \right)$ into Eq.~(\ref{ct_refless}) and simplifying, we get
%%%%%%%%%%%%%%%%%%%%  ct_dg   %%%%%%%%%%%%%%%%%%%%%%%%%%%%%%%%%
\begin{equation}\label{ct_dg}
{C_{dg}}\left( t \right) = \frac{V_{dc}}{{v_{dg}^ + \left( {t,0} \right) + {V_{dc}}}}\left[ {\frac{\sigma }{{{Z_{0}V_{dc}}}}{e^{ - \frac{{{{\left( {t - \mu } \right)}^2} - {\sigma ^2}}}{{2{\sigma ^2}}}}} + } {C_i} \right]
\end{equation}
%%%%%%%%%%%%%%%%%%%%%%%%%%%%%%%%%%%%%%%%%%%%%%%%%%%%%%%%%
which is the required modulation of the capacitor to eliminate the reflected pulse in the transmission line of Fig.~\ref{one} with $v^ + \left( {t,z} \right)=v_{dg}^ + \left( {t,z} \right)$. Fig.~\ref{two}(a) displays the temporal evolution of ${C_{dg}}$ for three different values of $V_{dc}$. One can observe that as the incoming pulse approaches the capacitor its capacitance (${C_{dg}}$) starts varying in time, as expected. During the transient time, the range of variation of ${C_{dg}}$ depends on $V_{dc}$. The larger $V_{dc}$ is, the smaller the range of variation of ${C_{dg}}$ is. After the transient time, ${C_{dg}}$ returns to its initial value ($C_i$) for all the three $V_{dc}$ values. This fact can also be rigorously seen by taking the limit $t \to \infty$ in Eq.~(\ref{ct_dg}). To understand the physical reason behind this result, we obtain the time-dependent equation of the charge in the capacitor by combining Eq.~(\ref{ct_dg}) with Eq.~(\ref{dqt})
%%%%%%%%%%%%%%%%%%%%  dqt_dg  %%%%%%%%%%%%%%%%%%%%%%%%%%%%%%%%
\begin{equation}\label{dqt_dg}
\Delta {q_{dg}}\left( t \right) = {\frac{\sigma }{{{Z_{0}}}}{e^{ - \frac{{{{\left( {t - \mu } \right)}^2} - {\sigma ^2}}}{{2{\sigma ^2}}}}} } 
\end{equation}
%%%%%%%%%%%%%%%%%%%%%%%%%%%%%%%%%%%%%%%%%%%%%%%%%%%%%%%%%
The temporal evolution of the charge is shown in Fig.~\ref{two}(b). $\Delta{q}_{dg}$ exhibits a Gaussian-like temporal variation and tends to zero for $t \to \infty$. As a result, the charge stored in the capacitor is identical before and after the transient time. And, in contrast to ${C_{dg}}$, $\Delta{q}_{dg}$ does not depend on $V_{dc}$. This behavior of $\Delta{q}_{dg}$ is expectedly consistent with the law of charge conservation, which, for the system under study, can be expressed as $\Delta {q_{dg}}\left( t \right) = \int\limits_{-\infty}^t i_{dg}^+\left( {\varepsilon ,0} \right)d\varepsilon $ with $i_{dg}^+=v_{dg}^+/Z_0$. As can be seen from the former equation, $\Delta{q_{dg}}$ depends only on the voltage of the pulse, but not on $V_{dc}$.
% $V_{dc}$ does not play any role because, as discussed above, it is assumed that the pulse starts propagating toward the capacitor when it has already been charged by the DC source, which remains then connected to the system for the entire process.%
 For large $t$ (after the transient time), given the odd symmetry of $v_{dg}^+$ [see the inset of Fig.~\ref{two}(a)], the previous integral is zero, which implies that the amount of current flowing in and out of the capacitor during the transient time is the same, resulting in a zero net flow of charge. Thus, the initial and final capacitance/charge values are identical. Therefore, for an incoming pulse with the temporal profile given by the derivative of a gaussian pulse, the capacitor of the system under study (Fig.~\ref{one}), with a capacitance variation as in Eq.~(\ref{ct_dg}), eliminates the reflected pulse and does not accumulate any additional charge with respect to the initial one provided by the DC source. This analysis also explains why the peak value of $C$ decreases with increasing $V_{dc}$ and can be generalized to any pulse with a zero net charge.

%allowing the system to return to the initial steady-state (before the pulse is injected into the transmission line)
%%%%%%%%%%%%%%%%%%%%%%%%%%%%%%%%%%%%%%%%%%%%%%%%%%%%%%%%%%%%%%%%%%%%%%%%%%%%%%%%%%%%%%%%%%
\subsection{Gaussian pulse} \label{subsec3B}
%%%%%%%%%%%%%%%%%%%%%%%%%%%%%%%%%%%%%%%%%%%%%%%%%%%%%%%%%%%%%%%%%%%%%%%%%%%%%%%%%%%%%%%%%%
Now, let us consider the following incoming Gaussian pulse: $v_g^ + \left( {t,z} \right) = A{e^{ - \frac{{{{\left( {t - {v_p}z - \mu } \right)}^2}}}{{2{\sigma ^2}}}}}$ with $\mu$ the position of the peak $A$, which it is assumed to be a real number, and $\sigma$, the standard deviation, controlling the temporal extension of the pulse. Substituting $v_g^ + \left( {t,z} \right)$ into Eq.~(\ref{ct_refless}), we obtain the required modulation of the capacitive load in the system under investigation (Fig.~\ref{one}) to achieve the reflectionless condition for an incoming gaussian pulse
%%%%%%%%%%%%%%%%%%%%  ct_g   %%%%%%%%%%%%%%%%%%%%%%%%%%%%%%%%%
\begin{equation}\label{ct_g}
C_{g}\left( t \right) = \frac{V_{dc}}{{A{e^{ - \frac{{{{\left( {t - \mu } \right)}^2}}}{{2{\sigma ^2}}}}} + {V_{dc}}}}\left[ {\frac{{A\sigma \sqrt {\frac{\pi }{2}} }}{{{Z_0} {{V_{dc}}} }}} 
{\left(1+ {{\rm{erf}}\left( {\frac{{t - \mu }}{{\sqrt 2 \sigma }}} \right)} \right) + C_i} \right]
\end{equation}
%%%%%%%%%%%%%%%%%%%%%%%%%%%%%%%%%%%%%%%%%%%%%%%%%%%%%%%%%
%%%%%%%%%%%%%%%%%%%%%%%%%%%%%%%%%%%%%%%%%%%%%%%%%%%%%%%%%
\begin{figure}[ht]
\centering
\includegraphics[width=0.85\columnwidth]{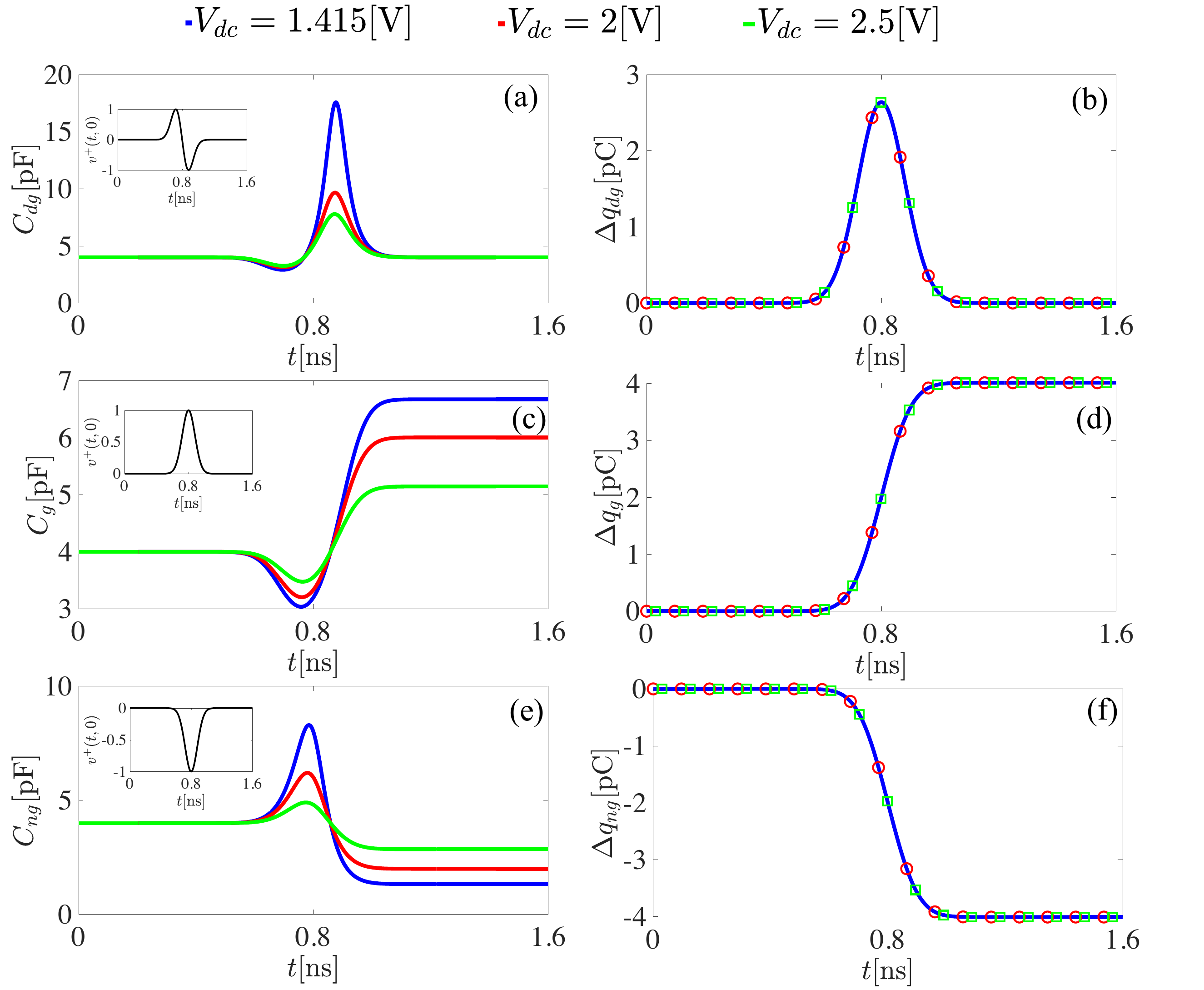}
\caption{Time dependence of the capacitance and charge experienced by the capacitor in the system of Fig.~\ref{one} to achieve the reflectionless condition for three different temporal shapes of the incoming pulse: (a) and (b) first-order derivative of the Gaussian function; (c) and (d) Gaussian function with positive peak: (e) and (f) Gaussian function with negative peak. Each panel display three curves corresponding to three different values of $V_{dc}$ indicated above the panels (a) and (b). The insets in (a), (c), and (d) show the profile of the considered incoming pulse. The results were obtained for $Z_0=50 \Omega$, $C_{i}=4$pF, $\mu=0.8$ns, and $\sigma=0.08$ns.} 
\label{two}
\end{figure}
%%%%%%%%%%%%%%%%%%%%%%%%%%%%%%%%%%%%%%%%%%%%%%%%%%%%%%%%%%%%%%%%%%%%%%%%%%%%%%%
\noindent with erf being the Gauss error function [$\rm{erf} \left(y\right) = \frac{2}{\sqrt{\pi}}\int\limits_{0}^y   e^{-y^2} dy$]. Combining the previous equation with Eq.~(\ref{dqt}), we obtain the time-dependent equation of the charge
%%%%%%%%%%%%%%%%%%%%  dqt_g  %%%%%%%%%%%%%%%%%%%%%%%%%%%%%%%%%
\begin{equation}\label{dqt_g}
\Delta{q}_{g}\left( t \right) = {\frac{{A\sigma \sqrt {\frac{\pi }{2}} }}{{{Z_0} }}} 
{\left(1+ {{\rm{erf}}\left( {\frac{{t - \mu }}{{\sqrt 2 \sigma }}} \right)} \right) }
\end{equation}
%%%%%%%%%%%%%%%%%%%%%%%%%%%%%%%%%%%%%%%%%%%%%%%%%%%%%%%%%
Eqs.~(\ref{ct_g}) and (\ref{dqt_g}), for large $t$, simplifies to 
%%%%%%%%%%%%%%%%%%%%  ct_g_inf  %%%%%%%%%%%%%%%%%%%%%%%%%%%%%%%%
\begin{equation}\label{ct_g_inf}
C_{g}\left( t\rightarrow{\infty} \right) =\sqrt{2\pi} \frac{A \sigma}{Z_0 V_{dc}}+C_i
\end{equation}
%%%%%%%%%%%%%%%%%%%%%%%%%%%%%%%%%%%%%%%%%%%%%%%%%%%%%%%%%
and
%%%%%%%%%%%%%%%%%%%%  dqt_g_inf  %%%%%%%%%%%%%%%%%%%%%%%%%%%%%%%
\begin{equation}\label{dqt_g_inf}
\Delta{q}_{g}\left( t\rightarrow{\infty} \right) = {\frac{{A\sigma \sqrt {2{\pi }} }}{{{Z_0} }}},
\end{equation}
%%%%%%%%%%%%%%%%%%%%%%%%%%%%%%%%%%%%%%%%%%%%%%%%%%%%%%%%%
respectively. From Eqs.~(\ref{ct_g}) and (\ref{dqt_g}), we can see that $C_{g}$ and $\Delta{q}_{g}$ may behave differently depending on the sign of the Gaussian pulse amplitude ($A$). Let us consider, first, the case with $A>0$, say $A=1$. With this amplitude of the gaussian pulse, the temporal evolution of $C_{g}\left( t \right)$ is shown in Fig.~\ref{two}(c) for three different $V_{dc}$ values. As observed, $C_{g}\left( t \right)$ varies during the transient time, and then stabilizes at a certain value, reaching a value that is higher than the initial one. The range of variation decreases with increasing $V_{dc}$. Thus, the value of $C_{g}\left( t \right)$ after the transient time (for large $t$) gets closer to the initial one for higher $V_{dc}$, as predicted by Eq.~(\ref{ct_g_inf}). According to Eq.~(\ref{dqt_g}), the temporal evolution of $\Delta{q}_{g}\left( t \right)$, which is shown in Fig.~\ref{two}(d), is independent of $V_{dc}$. And, after the transient time, the charge stored in the capacitor is increased. As can be seen in Eq.~(\ref{dqt_g_inf}), the amount of charge stored in the capacitor after the transient depends on the parameters characterizing the gaussian pulse, specifically the amplitude $A$ and the standard deviation $\sigma$. Note that the discussion on the law of charge conservation carried out for the incoming pulse considered in the previous section can be repeated for an incoming pulse with an arbitrary temporal shape. Accordingly, the growth of charge experienced by the capacitor for the incoming gaussian pulse with positive amplitude results from the current  ($\frac{{v_{g}^ + \left( {t ,z} \right)}}{{{Z_0}}}$) that this pulse induces in the transmission line, which implies a positive net flow of charge toward the capacitor. This charge is accumulated in the capacitor and is responsible for the different final capacitor value than the initial one. 

Now, let us consider the case when the amplitude of the incoming gaussian pulse [$v_g^ + \left( {t,z} \right)$] is a negative real number, $A<0$. The profile of $v_g^ + \left( {t,z} \right)$ with $A=-1$ is shown in the inset of Fig.~\ref{two}(e). To eliminate the reflection from this pulse the capacitor of the system in Fig.~\ref{one} needs to experience the temporal variation of its capacitance and charge shown in Figs.~\ref{two}(e) and (f), respectively. These plots are obtained from Eqs.~(\ref{ct_g}) and (\ref{dqt_g}) and, to differentiate the notation between these results and the ones of Figs.~\ref{two}(c) and (d) (corresponding to $A=1$), we replaced $C_{g}\left( t \right)$ and $\Delta {q_{g}}\left( t \right)$ with $C_{ng}\left( t \right)$ and $\Delta {q_{ng}}\left( t \right)$, respectively. As one can observe from Figs.~\ref{two}(e) and (f), the temporal evolution of $C_{ng}\left( t \right)$ and $\Delta {q_{ng}}\left( t \right)$ is flipped along the vertical axis with respect to the case with $A=1$ (see Figs.~\ref{two}(c) and (d)). With $A=-1$, or more generally when the amplitude of the gaussian pulse is negative, the current associated to this pulse flows toward the DC source. This current, according to the law of charge conservation, induces a flow of charge out of the capacitor. As a result, the charge stored in the capacitor and its capacitance after the transient decrease, as predicted by Eqs.~(\ref{ct_g_inf}) and (\ref{dqt_g_inf}) with $A<0$. Thus, when the peak of the gaussian pulse is negative, the capacitor experiencing a temporal modulation of its capacitance given by Eq.~(\ref{ct_g}) eliminates the reflected pulse and releases part of its charge into the transmission line.
%%%%%%%%%%%%%%%%%%%%%%%%%%%%%%%%%%%%%%%%%%%%%%%%%%%%%%%%%

%%%%%%%%%%%%%%%%%%%%%%%%%%%%%%%%%%%%%%%%%%%%%%%%%%%%%%%%%
\section{Energy exchange process} \label{sec4}
%%%%%%%%%%%%%%%%%%%%%%%%%%%%%%%%%%%%%%%%%%%%%%%%%%%%%%%%%
Another essential aspect of the system in Fig.~\ref{one} is the energy exchange process between the dynamic energy associated to the incoming pulse and the electrostatic energy stored in the capacitor before and after the transient time. %To this end, in this section, we investigate such an energy exchange process considering the difference between the energy in the system before and after the incoming voltage pulse is captured by the capacitor.%
Exploring this aspect will address the following fundamental question: How much of the incoming pulse's energy is trapped in the capacitor, and how much is transferred to the modulation source? The dynamic energy associated to the pulse is given by
%As the energy in the system before the pulse has reached the capacitor, we consider the electrostatic energy stored in the capacitor provided by the DC source (${W_i^s} = \frac{1}{2}{q_i}{V_{dc}}$) and the dynamic energy induced by the pulse, which is given by
%%%%%%%%%%%%%%%%%%%%  ene_pul  %%%%%%%%%%%%%%%%%%%%%%%%%%%%%%%%
\begin{equation}\label{ene_pul}
W_i^d = \int\limits_{-\infty}^\infty  {\left( {{v^ + }\left( {t,z} \right) + {V_{DC}}} \right){i^ + }\left( {t,z} \right)dt}=\int\limits_{-\infty}^\infty   {{v^ + }\left( {t,z} \right) {i^ + }\left( {t,z} \right)dt}+V_{dc}\int\limits_{-\infty}^\infty   {{i^ + }\left( {t,z} \right) dt}
\end{equation}
%%%%%%%%%%%%%%%%%%%%%%%%%%%%%%%%%%%%%%%%%%%%%%%%%%%%%%%%%
On the right-hand side of the previous equation, the first and second terms are the energy carried by the pulse ($W^p$) and the energy due to the current of a pulse flowing in a charged transmission line ($W^c$), respectively. The latter, upon applying the law of charge conservation and considering that the charge is conserved in time-varying capacitor, becomes $W^c = V_{dc}\Delta{q^{\infty}}=V_{dc}\left(q_f-q_i \right)$, with $q_f$ denoting the final charge stored in the capacitor. Note that $\Delta{q^{\infty}}$ coincides with Eq.~(\ref{dqt}) by letting $t\rightarrow{\infty}$. %Thus, the total energy in the system while the pulse is traveling in the transmission line can be expressed as $W_i^{tot} =W_i^s+W^p+W^c$. After the capacitor has captured the incoming pulse, we consider the energy stored in the capacitor for large $t$, which is simply given by ${W_f^{tot} } = \frac{1}{2}{q_f}{V_{dc}}$. Setting the energy balance as $\Delta W = W_f^{tot} - W_i^{tot} $, we get
Thus, Eq.~(\ref{ene_pul}) can be expressed in a compact form as $W_i^{d} =W^p+W^c$. Then, the energy balance is expressed as $\Delta W = W_f^{s}-W_i^d- W_i^{s} $, where ${W_i^s} = \frac{1}{2}{q_i}{V_{dc}}$ and ${W_f^{s} } = \frac{1}{2}{q_f}{V_{dc}}$ are the electrostatic energy stored in the capacitor before and after the transient time, respectively, $\Delta W$ is equal to the energy given by the modulation agent to the network. It is easy to show that
%%%%%%%%%%%%%%%%%%%%  ene_bal  %%%%%%%%%%%%%%%%%%%%%%%%%%%%%%%%
\begin{equation}\label{ene_bal}
\Delta W =  - W_{}^p- \frac{1}{2}{V_{dc}}\Delta q^{\infty} 
\end{equation}
%%%%%%%%%%%%%%%%%%%%%%%%%%%%%%%%%%%%%%%%%%%%%%%%%%%%%%%%%
We point out that such an energy-balance equation is general in the sense that is not restricted to a specific temporal shape of the incoming pulse and allows one to elucidate from first principles the energy exchange mechanism in the system of Fig.~\ref{one} operating in the reflectionless mode. Inspecting Eq.~(\ref{ene_bal}), one can observe that the first term on the right-hand side, which is related to the energy of the incoming pulse, is a negative term in the energy balance regardless of the pulse time shape and the DC source. This implies that a time-varying capacitor terminating a transmission line cannot trap the energy of the incoming pulse, which is transferred to the modulation source. On the other hand, the second term on the right-hand side of Eq.~(\ref{ene_bal}), which is related to the energy associated to the current of the pulse flowing in a charged transmission line ($W^c$), implies different energetic changes in the system depending on the sign of $\Delta q^{\infty}$. 

\begin{figure}[ht]
\centering
\includegraphics[width=0.7\columnwidth]{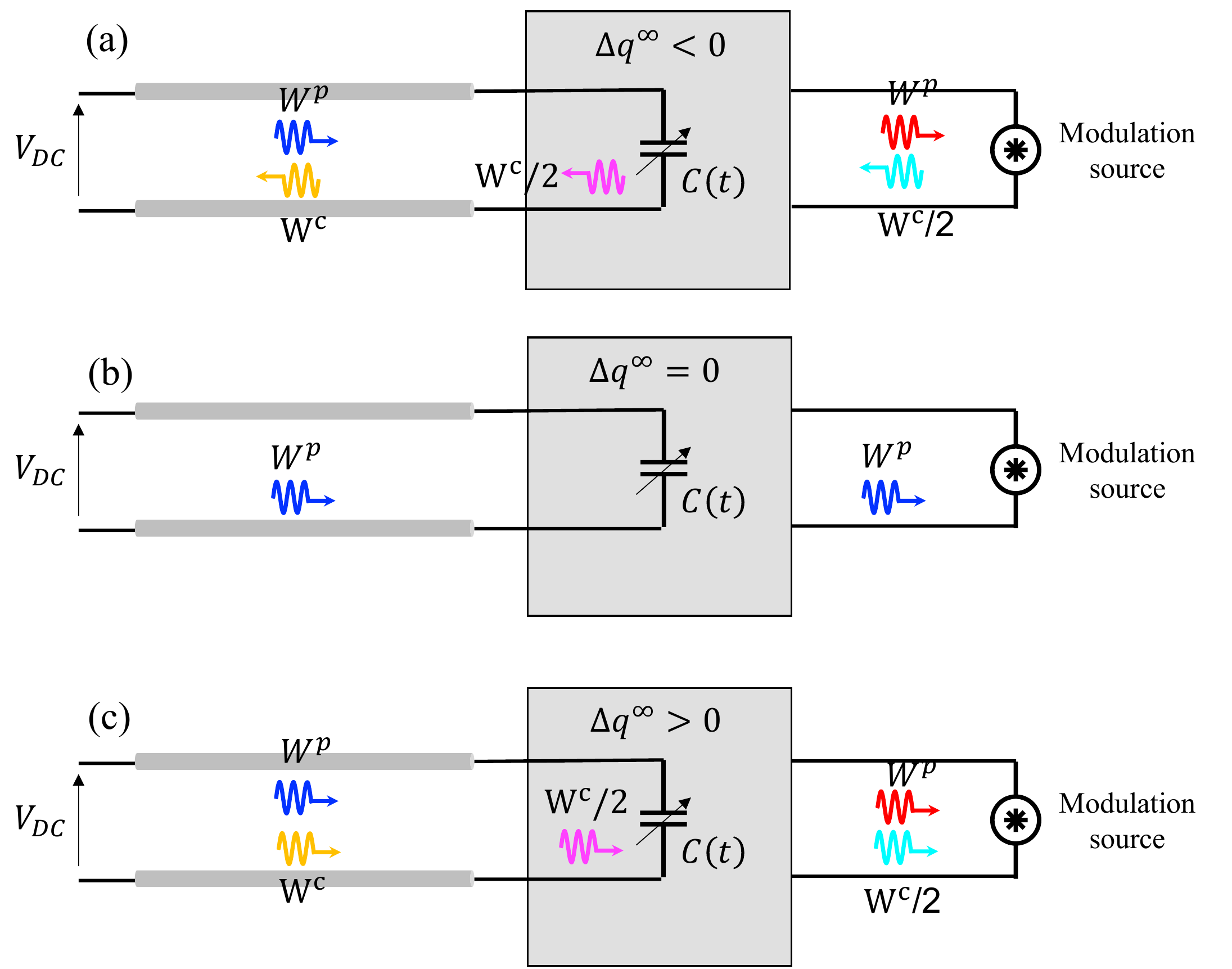}
\caption{Schematic representation of the energy exchange process experienced by the system in Fig.~\ref{one} operating in the reflectionless regime for incoming pulses that induce (a) a flow of charge moving away from capacitor ($\Delta q^{\infty}<0$), (b) a zero flow of charge ($\Delta q^{\infty}=0$), and (c) a flow of charge toward the capacitor ($\Delta q^{\infty}>0$).} 
\label{eb}
\end{figure}

For $\Delta q^{\infty}<0$, which occurs for incoming pulses such as the gaussian pulse with negative peak investigated above, the energy exchange process undergoing in the system is schematically shown in Fig.~\ref{eb}(a). As indicated in Sec.~\ref{sec2}, we assume $V_{dc}>0$. The following discussion can be easily extended to the case with $V_{dc}<0$. The energy associated to the current of the pulse flowing in a charged transmission line is a negative quantity ($W^c = V_{dc}\Delta{q^{\infty}}<0$) resulting in a depletion of the energy in the transmission of an amount equal to $W^c$ that ends up being accumulated in the DC source. As emerged from the second term on the right-hand side of Eq.~(\ref{ene_bal}), which is a positive quantity with $\Delta q^{\infty}<0$ and equals to $W^c$ except for a factor $1/2$, half of the energy in the transmission line is restored by the modulation source pumping energy into the system. The other half of the transmission line energy is restored by the capacitor losing part of its charge during the transient time ($\Delta q^{\infty}<0$). As discussed above and shown in Fig.~\ref{eb}(a), the energy of the incoming pulse is captured by the modulation source. %So far, we have discussed the energy exchange mechanisms undergoing among the different components of the system. Now, building on these energy exchanges, we elaborate on the system energy balance as a whole. From Eq.~(\ref{ene_bal}), one can observe that $\Delta W$ depends on $V_{dc}$ and as we are considering the case with $\Delta q^{\infty}<0$ we can identify different energy exchange regimes depending on the DC voltage across the capacitor.
With $\Delta q^{\infty}<0$, Eq.~(\ref{ene_bal}) results in three different energy exchange regimes depending on the DC voltage across the capacitor ($V_{dc}$). First, by setting $\Delta W=0$ in Eq.~(\ref{ene_bal}), we can find the DC voltage across the capacitor ($V_{dc}^{*}=2W^{p}/ \lvert{\Delta q^{\infty}}\rvert$) such that the energy of the system is ``conserved.'' As shown in Fig.~\ref{eb}(a), the modulation source pumps energy into system to restore the energy in the transmission line and captures the energy of the incoming pulse. These two energies balance out when $V_{dc}=V_{dc}^{*}$ and, as a result, $\Delta W=0$. This implies that the energy accumulated in the DC source and the energy captured by the modulation source are identical. When $V_{dc}<V_{dc}^{*}$, $\Delta W<0$ implying that the energy of the pulse transferred to the modulation source exceeds the energy that the modulation source pumps into the system and accumulates in the DC source. On the other hand, when $V_{dc}>V_{dc}^{*}$, the modulation source pumps into the system an amount of energy, which is accumulated in the DC source, that is higher the one received from the incoming pulse ($\Delta W>0$). 

For $\Delta q^{\infty}=0$, which indicates that the initial and final energy stored in the capacitor are identical, according to Eq.~(\ref{ene_bal}), the only energy exchange taking place in the system is the transfer of the pulse energy to the modulation source [see Fig.~\ref{eb}(b)]. This scenario occurs for the family of incoming pulses whose current induces a no net flow of charge along the transmission line, such as the pulse investigated in Sec.~\ref{subsec3A} and shown in the inset of Fig.\ref{two}(a). 
%%%%%%%%%%%%%%%%%%%%%%%%%%%%%%%%%%%%%%%%%%%%%%%%%%%%%%%%%
%%%%%%%%%%%%%%%%%%%%%%%%%%%%%%%%%%%%%%%%%%%%%%%%%%%%%%%%%

The last case that needs to be considered is when $\Delta q^{\infty}>0$, which is shown in Fig.~\ref{eb}(c). In this case, the energy associated to the current of the pulse flowing in a charged transmission line ($W^c = V_{dc}\Delta{q^{\infty}}>0$) travels toward the capacitor. By comparing $W^c$ with the expression of the difference between the initial and final energy stored in the capacitor ($ \frac{1}{2}{\Delta q^{\infty}}{V_{dc}}$), one can observe that they are equal except for a factor $1/2$. Hence, the capacitor is only able to capture half of $W^c$. The other half of $W^c$, in addition to the energy of the incoming pulse ($W^p$), is transferred to the modulation source, as emerged from the energy balance equation [Eq.~(\ref{ene_bal})] that results in $\Delta W<0$. %Increasing $V_{dc}$ implies higher $W^c$ but, as the capacitor captures only half of this energy, $\Delta W$ decreases linearly with respect to $V_{dc}>0$.%
For the set of incoming pulses that induce a net flow of charge moving toward the capacitor ($\Delta q^{\infty}>0$), such as the gaussian pulse with a positive peak investigated in Sec.~\ref{subsec3B}, the capacitor captures only half of the energy induced by the current of the pulse flowing in a charged transmission line. The other half of this energy and the pulse's energy are transferred to the modulation source.

In the first part of this section, we discussed the energy exchange process undergoing in the system under study (Fig.~\ref{one}) in general in the sense that has not been limited to incoming pulses with specific temporal evolution. Now, to further investigate such a process, including the temporal evolution of the energy balance, we focus on the incoming pulses investigated in Secs.~\ref{subsec3A} and \ref{subsec3B}.

%%%%%%%%%%%%%%%%%%%%%%%%%%%%%%%%%%%%%%%%%%%%%%%%%%%%%%%%%
\subsection{Energy balance for an incoming pulse with the temporal profile of the first-derivative of a gaussian function} \label{subsec4A}
\begin{figure}[ht]
\centering
\includegraphics[width=0.8\columnwidth]{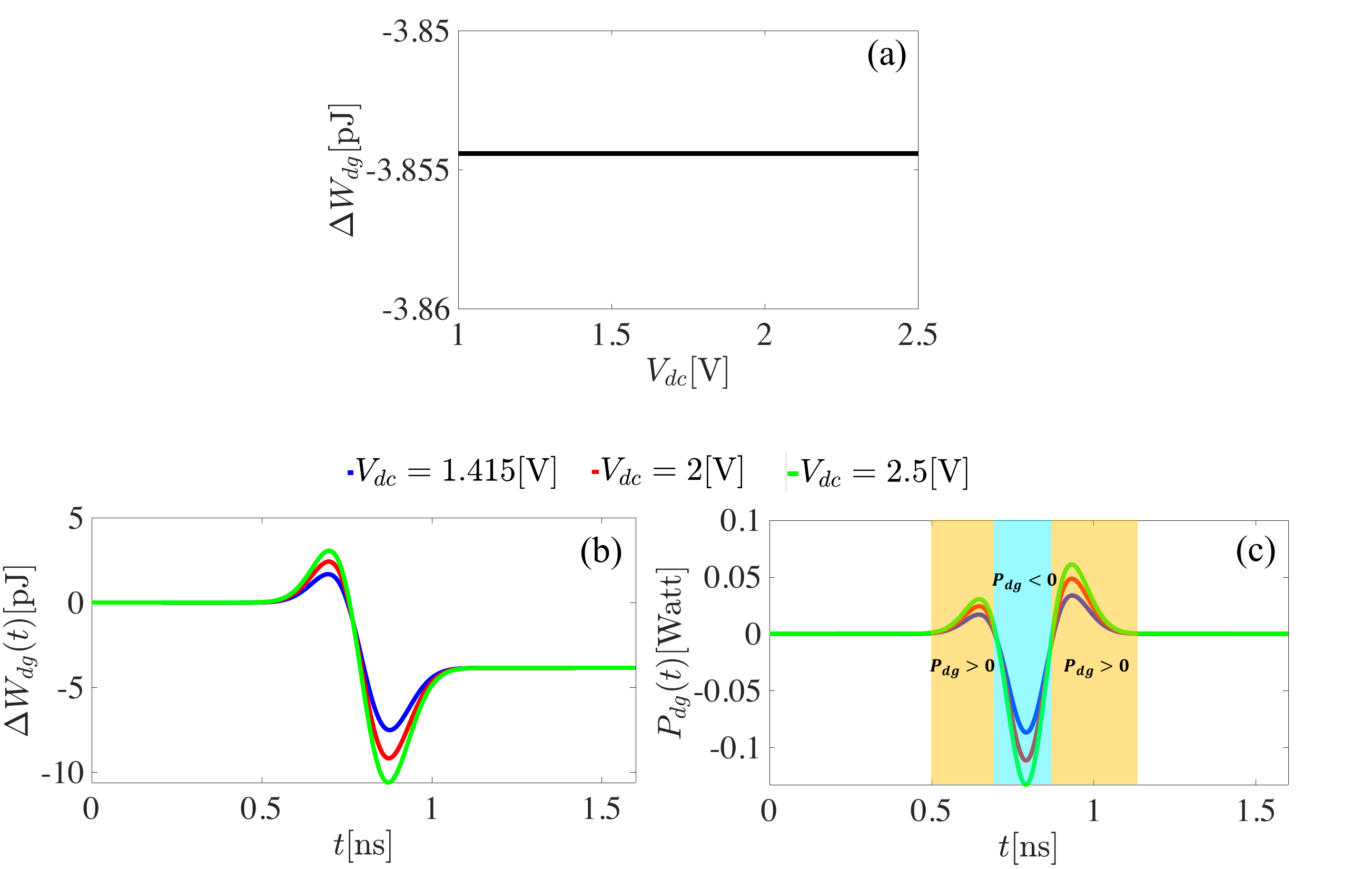}
\caption{Energy exchange process of the system in Fig.~\ref{one} operating in the reflectionless regime for an incoming pulse given by the first-order derivative of a gaussian function [$v_{dg}^ + \left( {t,z} \right)$]. (a) Energy balance as a function of the DC voltage across the capacitor. (b) Instantaneous energy balance and (c) its temporal derivative for three different values of $V_{dc}$, which are indicated above these panels. The results were obtained for $Z_0=50 \Omega$, $C_{i}=4$pF, $\mu=0.8$ns, and $\sigma=0.08$ns.} 
\label{three}
\end{figure}

%%%%%%%%%%%%%%%%%%%%%%%%%%%%%%%%%%%%%%%%%%%%%%%%%%%%%%%%%
As can be seen in Fig.~\ref{two}(b), for an incoming pulse consisting of the first-order derivative of the gaussian function [$v_{dg}^ + \left( {t ,z} \right)$], the final charge in the capacitor (after the transient) is identical to the initial one. When the capacitor accumulates no additional charge during the transient ($\Delta q^{\infty}=0$), the energy-balance equation [Eq.~(\ref{ene_bal})] becomes 
%%%%%%%%%%%%%%%%%%%%  ene_bal_dg %%%%%%%%%%%%%%%%%%%%%%%%%%%%%%
\begin{equation}\label{ene_bal_dg}
\Delta W_{dg} = - W^p_{dg}
\end{equation}
%%%%%%%%%%%%%%%%%%%%%%%%%%%%%%%%%%%%%%%%%%%%%%%%%%%%%%%%%
with $W^p_{dg}$ the energy carried by $v_{dg}^ + \left( {t ,z} \right)$, which is transferred to the modulation source, as discussed in the previous section. Since $\Delta W_{dg}$ is independent of $V_{dc}$ [see Fig.~\ref{three}(a)], to get a better understanding of the role played by the DC voltage source from energetic standpoint, in Fig.~\ref{three}(b), we have plotted the temporal evolution of the energy balance, which can be expressed as
%%%%%%%%%%%%%%%%%%%%  inst_ene_bal_dg %%%%%%%%%%%%%%%%%%%%%%%%%%%%
\begin{equation}\label{inst_ene_bal_dg}
\Delta W_{dg}\left(t\right) = \frac{1}{2}\left(v_{dg}^ +\left( {t ,z} \right)+V_{dc}\right) v_{dg}^ +\left( {t ,z} \right) C_{dg}\left(t\right)- \frac{1}{2}{V_{dc}}\Delta q_{dg}\left(t\right)  - \int\limits_{-\infty}^t {\frac{\left({v_{dg}^ + \left( {\varepsilon ,z} \right)}\right)^2}{{{Z_0}}}d\varepsilon }
\end{equation}
%%%%%%%%%%%%%%%%%%%%%%%%%%%%%%%%%%%%%%%%%%%%%%%%%%%%%%%%%
One can observe that, during the transient time, $\Delta W_{dg}\left(t\right)$ depends on $V_{dc}$ exhibiting a larger swing for higher $V_{dc}$ values. This behavior seems to be related to the fact that more energy is stored in the capacitor for higher $V_{dc}$ values. And while the capacitance is experiencing a temporal modulation, more energy is exchanged between the capacitor and the modulation source. After the transient time, the three curves expectedly converge to the same constant negative quantity ($- W^p_{dg}$). We can further investigate this energy exchange mechanism by looking at the instantaneous power balance ($P_{dg}$), obtained as the derivative of $\Delta W_{dg}\left( t \right)$ with respect to time. The temporal evolution of $P_{dg}$ for three values of $V_{dc}$ is shown in Fig.~\ref{three}(c). When $P_{dg}$ is positive (see orangish filled boxes in Fig.~\ref{three}(c)), the instantaneous energy of the pulse is lower than the energy required to modulate the capacitor, which, assuming a mechanical capacitor, is given by $W_{mech}\left(t\right) = \frac{1}{2}\left(v_{dg}^ +\left( {t ,z} \right)+V_{DC}\right)^2 C_{dg}\left(t\right)$, and the modulation source needs to supply energy into the system. On the other hand, when $P_{dg}$ is negative (see cyanish filled box in Fig.~\ref{three}(c)), the instantaneous energy of the pulse exceeds the energy required to modulate the capacitor, and a portion of it ends up to the modulation source. %To summarize the results presented in this subsection, for incoming pulses whose current does not involve any net flow of charge, such as the derivative of the Gaussian pulse, no energy is accumulated or stored in the capacitor. The energy of the incoming pulse is completely transferred to the modulation source regardless of the $V_{dc}$ values.
%%%%%%%%%%%%%%%%%%%%%%%%%%%%%%%%%%%%%%%%%%%%%%%%%%%%%%%%%
%%%%%%%%%%%%%%%%%%%%%%%%%%%%%%%%%%%%%%%%%%%%%%%%%%%%%%%%%
%%%%%%%%%%%%%%%%%%%%%%%%%%%%%%%%%%%%%%%%%%%%%%%%%%%%%%%%%
\subsection{Energy balance for an incoming pulse with the temporal profile of a Gaussian function} \label{subsec4B}
\begin{figure}[ht]
\centering
\includegraphics[width=0.8\columnwidth]{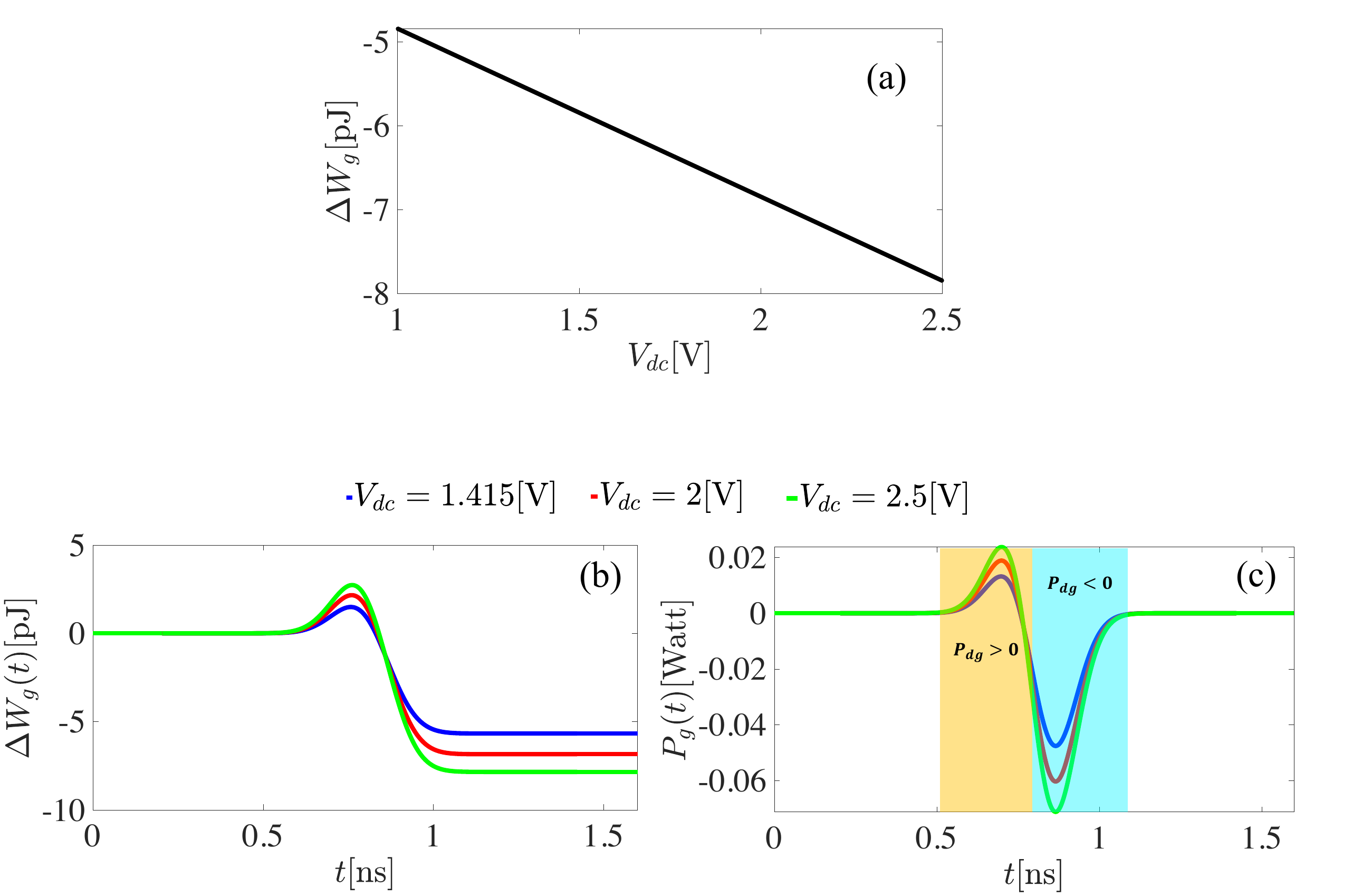}
\caption{Energy exchange process of the system in Fig.~\ref{one} for an incoming pulse given by a Gaussian function [$v_{g}^ + \left( {t,z} \right)$] with the peak $A=1$. (a) Energy balance as a function of the DC voltage across the capacitor. (b) Instantaneous energy balance and (c) its temporal derivative for three different values of $V_{dc}$, which are indicated above these panels. The results were obtained for $Z_0=50 \Omega$, $C_{i}=4$pF, $\mu=0.8$ns, and $\sigma=0.08$ns.} 
\label{four}
\end{figure}

%%%%%%%%%%%%%%%%%%%%%%%%%%%%%%%%%%%%%%%%%%%%%%%%%%%%%%%%%
Combining Eq.~(\ref{dqt_g_inf}) with Eq.~(\ref{ene_bal}), we obtain the energy-balance equation for the incoming gaussian pulse $v_{g}^ + \left( {t,z} \right)$ [Sec.~\ref{subsec3B}]
%%%%%%%%%%%%%%%%%%%%  ene_bal_g %%%%%%%%%%%%%%%%%%%%%%%%%%%%%%%
\begin{equation}\label{ene_bal_g}
\Delta W_g =  - {\frac{{A\sigma }}{{{Z_0} }}}  \sqrt {\frac{\pi }{2}} V_{dc}- W_{g}^p
\end{equation}
%%%%%%%%%%%%%%%%%%%%%%%%%%%%%%%%%%%%%%%%%%%%%%%%%%%%%%%%%
with $W_{g}^p$ representing its energy. One can observe that this energy-balance equation may involve different energetic considerations depending on the sign of $A$ (the peak of the Gaussian pulse). First, let us assume $A>0$, say $A=1$ for the sake of consistency with the results shown in Figs.~\ref{two}(c) and (d).  With an incoming Gaussian pulse with this peak, which induces an increase of the final charge stored in the capacitor [see Fig.~\ref{two}(d)], the system experiences the energy exchange mechanism discussed in Sec.~\ref{sec4} for the general class of incoming pulses with $\Delta q^{\infty}>0$ [see Fig.~\ref{eb}(c)]. The energy of the pulse ($W_{g}^p$) and half of the energy associated to the current of the pulse flowing in a charged transmission are transferred to the modulation source, resulting in $\Delta W_g<0$ that decreases linearly with $V_{dc}$, as shown in Fig.~\ref{four}(a). The temporal evolution of $\Delta W_g$, which has been obtained by replacing in Eq.~(\ref{inst_ene_bal_dg}) $v_{dg}^ + \left( {t,z} \right)$, $C_{dg} \left( {t} \right)$, and $\Delta q_{dg} \left( {t} \right)$ with $v_{g}^ + \left( {t,z} \right)$, $C_{g} \left( {t} \right)$, and $\Delta q_{g} \left( {t} \right)$, respectively, is shown in Fig.~\ref{four}(b). As observed, higher $V_{dc}$ implies that more energy is exchanged between the capacitor and the modulation source, resulting in a large amount of energy transferred the modulation source, as predicted by Eq.~(\ref{ene_bal_g}). Fig.~\ref{four}(c) displays the instantaneous power balance, $P_g\left(t\right)$. During the transient time the modulation source, first, supplies power to the system ($P_g\left(t\right)>0$) and, then takes power from the system ($P_g\left(t\right)<0$). Overall, the power taken from the modulation source exceeds the power gain by the system, resulting in a flow power moving toward the modulation source, as expected. 

%%%%%%%%%%%%%%%%%%%%%%%%%%%%%%%%%%%%%%%%%%%%%%%%%%%%%%%%%
%%%%%%%%%%%%%%%%%%%%%%%%%%%%%%%%%%%%%%%%%%%%%%%%%%%%%%%%%

\begin{figure}[H]
\centering
\includegraphics[width=0.8\columnwidth]{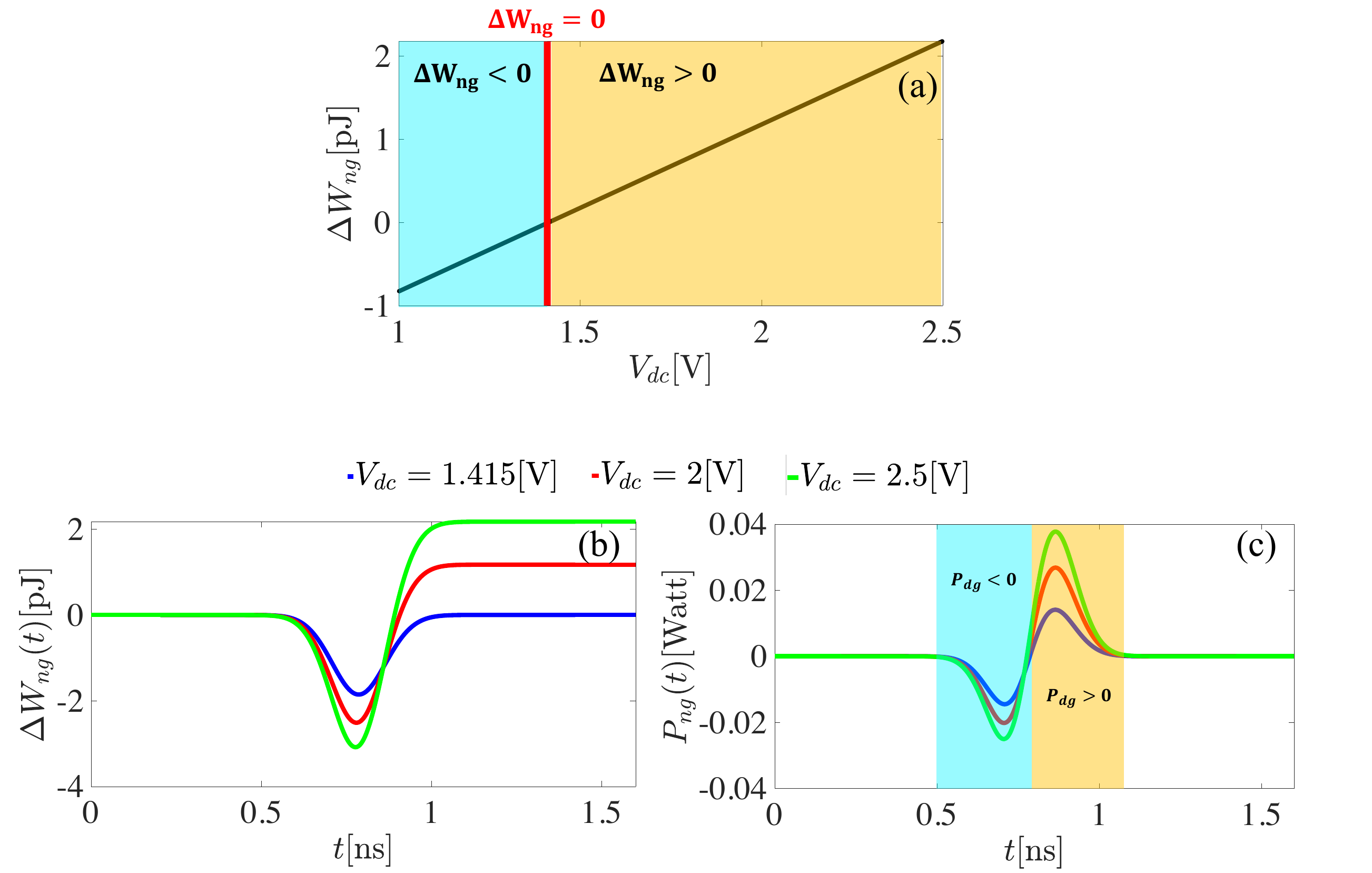}
\caption{Energy exchange process of the system in Fig.~\ref{one} for an incoming pulse given by a Gaussian function [$v_{ng}^ + \left( {t,z} \right)$] with the peak $A=-1$. (a) Energy balance as a function of the DC voltage across the capacitor. (b) Instantaneous energy balance and (c) its temporal derivative for three different values of $V_{dc}$, which are indicated above these panels. The results were obtained for $Z_0=50 \Omega$, $C_{i}=4$pF, $\mu=0.8$ns, and $\sigma=0.08$ns.} 
\label{five}
\end{figure}

Now, let us consider the case when the peak of the gaussian pulse is a negative real number, say $A=-1$ for consistency with the results in Figs.~\ref{two}(e) and (f). With this incoming pulse, which induces a decrease of the charge stored in the capacitor (see Fig.~\ref{two}(f)), the system experiences the energy exchange mechanism discussed in Sec.~\ref{sec4} for the general class of incoming pulses with $\Delta q^{\infty}<0$ [see Fig.~\ref{eb}(a)] and there are three different energetic regimes depending upon the DC voltage across the capacitor, as can be seen in Fig.~\ref{five}(a). Note that to differentiate this result to the one in Fig.~\ref{four}(a), as both are obtained from Eq.~(\ref{ene_bal}), we replaced $\Delta W_{g}$ with $\Delta W_{ng}$. In Fig.~\ref{five}(a), one can observe that there is a specific value of the DC voltage ($V_{dc}^{*}=1.415$[V]) for which $\Delta W_{ng}=0$, meaning that the system transfers and receives the same amount of energy from the modulation source. To better understand this energetic regime, we can look at the blue-solid curves in Figs.~\ref{five}(b) and (c) showing the temporal evolution of the energy balance [$\Delta W_{ng}\left(t\right)$] and power balance [$P_{ng}\left(t\right)$], respectively, for $V_{dc}=V_{dc}^{*}$. $\Delta W_{ng}\left(t\right)$ converging to zero after the transient indicates that the amount of instantaneous energy flowing from the modulation source to the system and accumulate in the DC source is equal to the energy of the pulse captured by the modulation source. This is also confirmed from the temporal profile of $P_{ng}\left(t\right)$. During the transient, the portion of the curve with $P_{ng}\left(t\right)<0$ is the upside down image of that with $P_{ng}\left(t\right)>0$. This symmetry implies that systems loses and gains the same amount of power, resulting in a power balance equal to zero as a whole. When the DC voltage across the capacitor is larger than $V_{dc}^{*}$, $\Delta W_{ng}>0$ (see orangish filled box in Fig.~\ref{five}(a)). The amount of energy pumped into the system by the modulation source that accumulates in the DC source exceeds the energy of the pulse captured by the modulation source. Figs.~\ref{five}(b) and (c) show the temporal evolution of $\Delta W_{ng}\left(t\right)$ and $P_{ng}\left(t\right)$, respectively, for two different values of $V_{dc}$ larger than $V_{dc}^{*}$ (red- and green-solid curves). $\Delta W_{ng}\left(t\right)$ exhibits higher swing for larger $V_{dc}$ resulting in more energy accumulated by the DC source. Similar physical consideration can be made from the temporal profile of $P_{ng}\left(t\right)$, which deviates from its symmetric behavior observed for $V_{dc}=V_{dc}^{*}$. The portion of the curve with $P_{ng}\left(t\right)>0$ increases faster than the one with $P_{ng}\left(t\right)<0$, indicating that the modulation source pumps into the system more energy than the one that it receives from the incoming pulse. Finally, for $V_{dc}<V_{dc}^{*}$, $\Delta W_{ng}<0$ (see cyanish filled box in Fig.~\ref{five}(a)), which is analogous to the cases of Figs.~\ref{three}(c) and \ref{four}(c). The modulation source pumps less energy into the system than it captures from the incoming pulse.
%%%%%%%%%%%%%%%%%%%%%%%%%%%%%%%%%%%%%%%%%%%%%%%%%%%%%%%%%
%%%%%%%%%%%%%%%%%%%%%%%%%%%%%%%%%%%%%%%%%%%%%%%%%%%%%%%%%

%%%%%%%%%%%%%%%%%%%%%%%%%%%%%%%%%%%%%%%%%%%%%%%%%%%%%%%%%
\section{Conclusion} \label{sec5}
%%%%%%%%%%%%%%%%%%%%%%%%%%%%%%%%%%%%%%%%%%%%%%%%%%%%%%%%%
In this paper, we theoretically investigate how a transmission line terminated with a capacitor actively modulated in time can eliminate a reflected signal. We show that when the capacitor is charged with a DC voltage source, the required temporal variation of its capacitance, which is derived in analytical form, avoids extreme and negative values. We elaborate on the temporal variation of the capacitance and the resulting temporal evolution of the charge in the capacitor for two illustrative incoming pulses. We derive from first principles the energy balance of such a reflectionless time-varying capacitor, showing that the energy of the incoming pulse, regardless of its temporal shape, is transferred to the modulation source. The energy balance is additionally investigated by considering two illustrative examples of incoming pulses.

\section*{Acknowledgment:} 
This work was supported by the National Science Foundation through the Industry-University Cooperative Research Centers (IUCRC) Center for Metamaterials under Grant 1624572. The work of David Gonz\'alez-Ovejero and Mario Junior Mencagli was supported by the Thomas Jefferson Fund of the Embassy of France in the United States and the FACE Foundation.\\
The authors Dimitrios L. Sounas and Mario Junior Mencagli thank Nader Engheta from the University of Pennsylvania for the fruitful discussion on this topic.
%\nocite{*}
\newpage
\bibliography{aapmsamp}% Produces the bibliography via BibTeX.

\end{document}